\documentclass[a4paper]{article} 
\usepackage[T1]{fontenc} 
\usepackage[utf8]{inputenc} 
\usepackage{ISMIRlbd, cite, times, amsmath, url}
\usepackage{hyperref}
\usepackage{url}
\usepackage{float}
\usepackage{graphicx}
\usepackage{booktabs}
\usepackage{color}
\usepackage{float}
\usepackage{subcaption}
\usepackage{cleveref}

\title{Automated Time-frequency Domain Audio Crossfades using Graph Cuts}

\twoauthors
 {Kyle Robinson}  {University of Waterloo \\ {\tt kyle.robinson@uwaterloo.ca}}
 {Dan Brown} {University of Waterloo}

\begin{document}

\maketitle

\begin{figure}[H]
    \centering
    \vspace{-3.8em}
    \includegraphics[width=\linewidth]{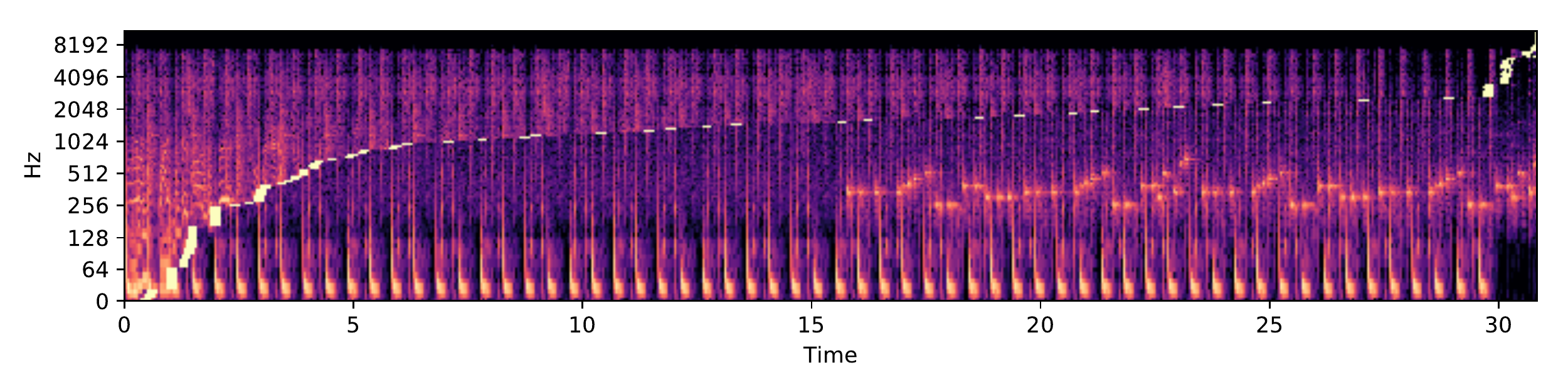}
    \vspace{-2.5em}
    \caption{This spectrogram shows overlapped segments of two music tracks after being combined and reconstructed along a per-frequency seam (bright yellow). The tracks were beat and tempo matched, then overlapped by 64 beats.}
    \label{fig:seam}
\end{figure}

\begin{abstract}
The problem of transitioning smoothly from one audio clip to another arises in many music consumption scenarios; especially as music consumption has moved from professionally curated and live-streamed radios to personal playback devices and services. Classically, transitioning from one song to another has been reliant on either pre-mixed transitions on recorded digital or physical media, hardware or software crossfading on the playback device, or professional transitions by a host or disk jockey (DJ). While options for software crossfading are ubiquitous on music streaming platforms and media players alike, these transitions pale in quality when compared to those manually applied by an audio engineer or DJ who can harmonically and rhythmically align tracks---and importantly---manually apply equalizer (EQ) filters during transitions. The application of EQ filters specifically allow for different transitions in different audio spectrums. For example, the bass register of one track can be made to replace the bass register of another track before transitioning the higher frequencies. Typically the task of deciding how and where to apply transitions in the frequency domain has been completed manually using a limited number of EQ filters.

There is much research on creating, sorting, and extending playlists so as to have tracks naturally flow into each other, as well as on determining optimal times to transition between similar tracks~\cite{Bittner,Parera2016,Gebhardt2016,Hirai2015,Lin2009}. Both of these research areas play a key role in synthesising a human DJ. To our knowledge, however, all of these approaches still rely on classical methods of transitioning tracks using amplitude in the time domain (crossfading).

Through the application of an existing visual texture extension algorithm borrowed from computer vision, we present the first steps toward a new method of automatically transitioning from one audio clip to another by discretizing the frequency spectrum into bins and then finding transition times for each bin.~\cite{Kwatra2003}.

We begin by phrasing the problem of transitioning from one song to another as a graph optimization problem:  the
graph represents the two songs in the transition range, and a cut happens when we transition from one song to the other at a particular
time point. To obtain these representations we first apply a short term fourier transform (STFT) to each song, and then convert the resulting complex time-frequency mapped amplitude values into real decibel values. In order to align the tracks we apply rudimentary tempo matching and beat alignment using the libROSA Python library, and overlap the tracks by a number of beats~\cite{Mcfee2015}. We call the resulting STFT transformed data of the first and second song's overlapped segments matrix \(A\) and matrix \(B\) respectfully, as seen in~\Cref{fig:AB}. Next, we define a simple loss function: 
\begin{equation}
    \label{eqn:weight}
    w^{k,l}_{i,j}(A,B) = ||A_{i,j}-B_{i,j}||+||A_{k,l}-B_{k,l}||
\end{equation}

We apply the loss function to each adjacent time-frequency bin and use the resulting values to assign weights to edges in an undirected graph with dimensions equal to \(A\) and \(B\). Finally, the resulting graphs left-most nodes are anchored to a source node, and the right-most nodes are anchored to a sink node. \Cref{fig:graph} shows a representation of the completed flow graph. In order to obtain a min-cut, we apply the Boykov-Kolmogorov algorithm~\cite{Boykov2004}. The indices of this min-cut are the seam where each frequency bin transitions.

In order to apply the found transition to the audio tracks, we concatenate the complex time-frequency song representations found earlier along the seam, and apply an inverse STFT to obtain the final audio transition seen in~\Cref{fig:seam}.

The work here presents an initial foray into automatically transitioning between songs in the time-frequency domain. The loss function described in \ref{eqn:weight} does not well characterize the inherent qualities found in music, but there is good reason to believe such a cost function can be found through further development. On harmonically similar tracks with similar tempi, the current implementation produces acoustically pleasing results.

\begin{figure}[t!]
    \centering
    \begin{minipage}[t][][t]{0.45\linewidth}
        \centering
        \includegraphics[width=\textwidth]{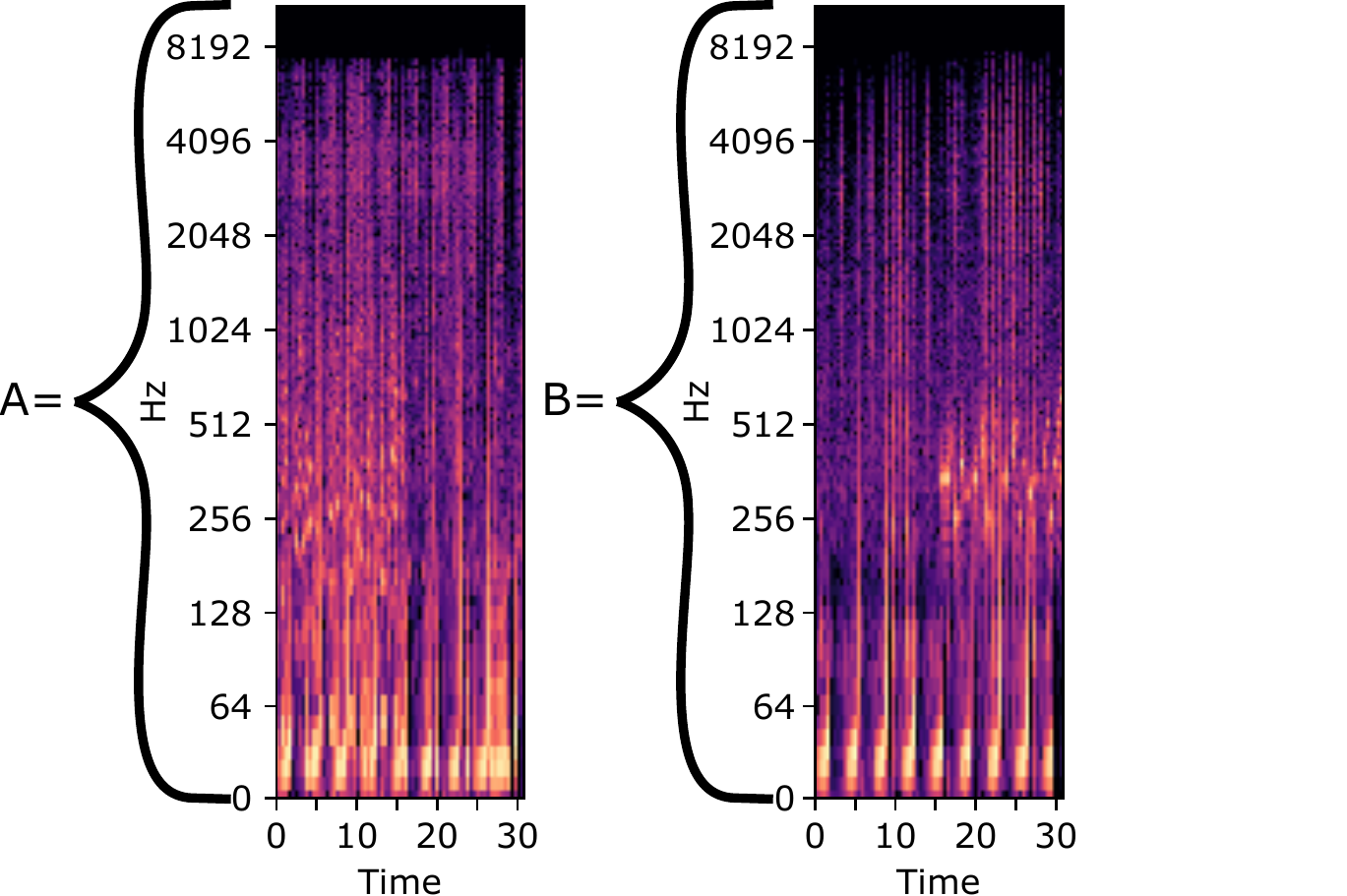}
        \caption{Overlapping song segment spectrogram's used to compute graph weights using \cref{eqn:weight}. Each segment represents 64 beats.}
        \label{fig:AB}
    \end{minipage}
    \hspace{0.075\linewidth}
    \begin{minipage}[t][][t]{0.4\linewidth}
        \centering
        \includegraphics[width=\textwidth]{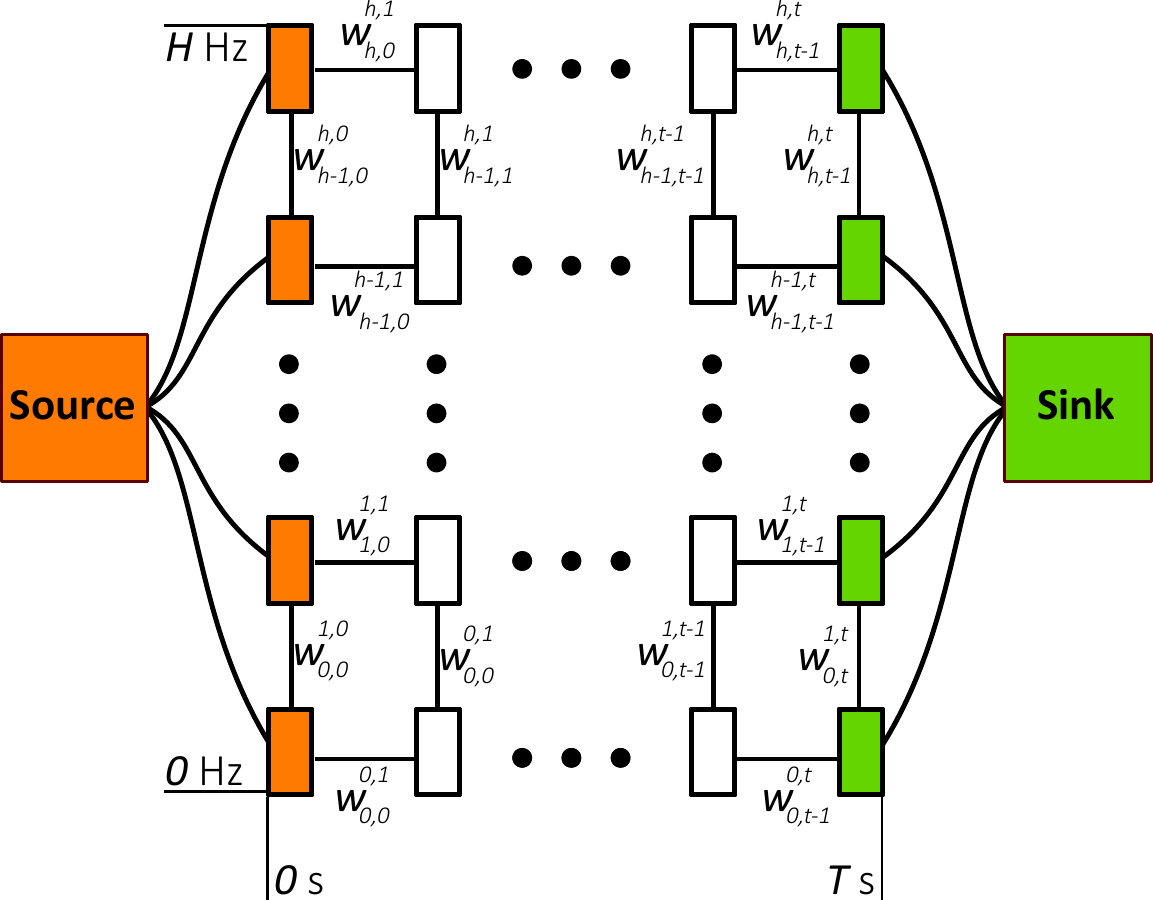}
        \caption{Flow graph model showing adjacent weight calculations. The orange and green nodes are fixed to the first and second songs respectfully. Nodes represent time-frequency bins.}
        \label{fig:graph}
    \end{minipage}
\end{figure}

\end{abstract}



{\footnotesize\bibliography{ISMIRtemplate}}

\end{document}